\documentclass[10pt]{article}

\usepackage{graphics}
\usepackage{graphicx}

\usepackage[a4paper, left=35mm,right=35mm,top=34mm,bottom=34mm]{geometry}
\usepackage[utf8]{inputenc}
\usepackage[T1]{fontenc}
\usepackage[english]{babel}

\usepackage{enumerate}
\usepackage{graphicx}
\usepackage{hyperref}
\hypersetup{
    colorlinks=true,
    linkcolor=blue,
    filecolor=magenta,      
    urlcolor=cyan,
}

\usepackage{mathtools,amsthm,amssymb,amsfonts}
\usepackage{algorithm}
\usepackage{algorithmic}
\makeatother
\theoremstyle{plain}

\theoremstyle{definition}

\theoremstyle{remark}

\usepackage{caption} 
\usepackage{listings}
\usepackage{color}
\definecolor{dkgreen}{rgb}{0,0.6,0}
\definecolor{gray}{rgb}{0.5,0.5,0.5}
\definecolor{mauve}{rgb}{0.58,0,0.82}
\usepackage{fancyhdr}

\author{
  {\normalsize Fr\'ed\'eric Magoul\`es}\thanks{CentraleSup\'elec, Universit\'e Paris-Saclay, France.
    (correspondence, frederic.magoules@hotmail.com).}
  \and
  {\normalsize Qinmeng Zou}\thanks{CentraleSup\'elec, Universit\'e Paris-Saclay, France.}
}
\title{A Novel Contactless Human Machine Interface based on Machine Learning}
\date{}

\begin{document}
\maketitle
\thispagestyle{fancy}

\begin{abstract}
\noindent This paper describes a global framework that enables contactless human machine interaction using computer vision and machine learning techniques.
The main originality of our framework is that only a very simple image acquisition device, as a computer camera, is sufficient to establish a rich human machine interaction as traditional devices such as mouse or keyboard.
This framework is based on well known computer vision techniques and efficient machine learning techniques are used to detect and track user hand gestures so the end user can control his computer using virtual interfaces with very simple gestures.
\end{abstract}


\section{Introduction}

Nowadays, the development of ubiquitous and embedded computing systems in our daily life implies new major challenges in human computer interface design~\cite{dix2004hci}.
In particular, usual interfaces such as keyboards and mouses are not suited for efficient and convenient communications with such systems.
As a consequence, many research efforts are carried to establish a more natural and convenient human computer communication without specific devices.
Good reviews on human machine interaction based on computer vision can be found in~\cite{cipolla1998cvh,jr99survey,moeslund06}.
Futuristic interfaces, as the famous interface presented in the 2002 science fiction film directed by Steven Spielberg entitled {\em Minority Report}, have inspired a lot of real advanced interfaces.
Most of these futuristic interfaces enable human computer communication at a distance without physical contact but they need specific devices such as hand gloves~\cite{sturman1994sgb} or other tracking devices for human motion capture.
Although these devices enable an accurate acquisition of the 3D motion, they are really expensive and cumbersome for real applications.

According to vision based human motion analysis~\cite{moeslund06,Poppe662:2007,wu99visionbased} or more particularly on vision based hand gesture interfaces~\cite{len02}, the different approaches can be classified according to different criteria:
(i) the acquisition device (mono~\cite{GiDa2005.1} or multi-camera~\cite{utsumi96} acquisition system),
(ii) the gesture representation and tracking approach (articulated kinematic or shape human model~\cite{ouhaddi-hand}, statistical shape based model~\cite{cootes95}, appearance-based approach~\cite{martin97}, and many other gesture modeling approaches),
(iii) the nature of the gesture to recognize (conversational gestures, controlling gestures, manipulative gestures or communicative gestures).
Our approaches is voluntarily a simple monocular vision based approach.
The design of our framework was made in respect with the following requirements:
(i) a single basic camera is used for the computer human interaction (a major choice for the accessibility of the framework in terms of price and infrastructure deployment),
(ii) a user friendly reconfigurable interface which is of prime importance for vision based user interface~\cite{icvs03},
(iii) a platform and hardware independent framework,
(iv) a real time video processing for a rich interaction, i.e., a small latency between the commands given by the end user with hand motions and the execution of the actions on the machine.

The plan of the paper is the following.
Section~\ref{sec:2} gives a global description of the framework and of its modular architecture.
In Section~\ref{sec:3}, the different modules of the framework are detailed together with some implementation issues.
Section~\ref{sec:4} presents some experimental results.
Finally, Section~\ref{sec:5} conclude this paper.

\section{Overview of the Framework}
\label{sec:2}

\begin{figure}
  \centering
  \includegraphics[scale=0.5]{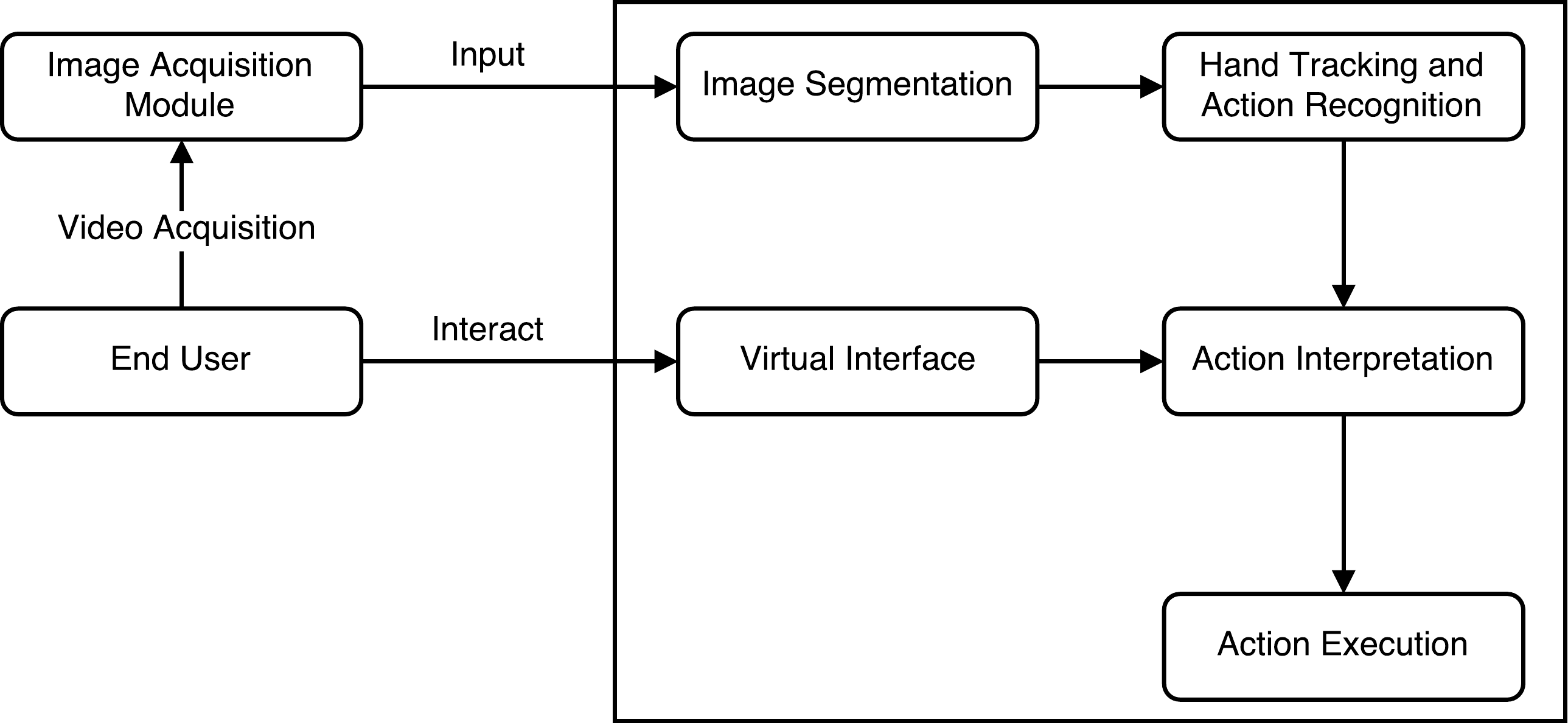}
  \caption{Global overview of the framework.}
  \label{fig:overview}
\end{figure}

This section presents a global description of our modular framework for the design of contactless human machine interfaces.
The different modules correspond to the different tasks needed to build a contactless human machine interface based on computer vision, i.e., 
(i) a video acquisition module for the capture of hand motions,
(ii) a real time video segmentation module for the segmentation of human hand gestures from the video frames,
(iii) a hand gesture tracking and recognition module,
(iv) an interface module for the communication between the end user and the machine,
(v) an engine module to execute the different actions requested by the user via the virtual interface.
A global overview of the framework and the orchestration of the different functionalities is described in Figure~\ref{fig:overview}.

\begin{figure}
  \centering
  \includegraphics[scale=0.4]{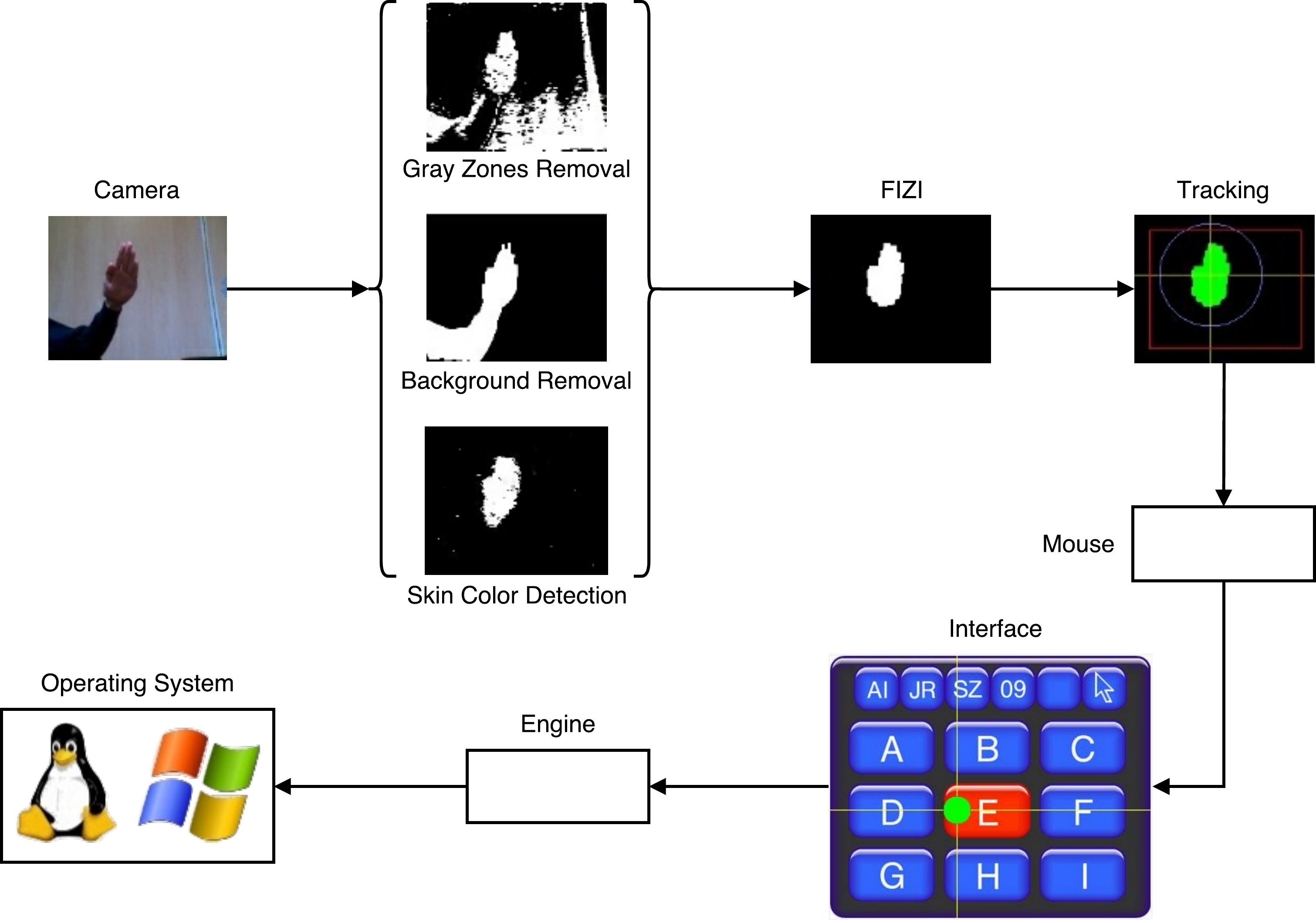}
  \caption{Architecture overview.}
  \label{fig:diagramarchitecture}
\end{figure}

The first module is FIZI module.
This module transforms each image captured by the camera into a mask of the zones of interest.
These connected zones of interest are the skin zones of the end user.
Then a selection step enables to select one region corresponding to one hand which is then tracked on all the video sequence.
The result of the Tracking module is then sent to Mouse module which maps it to a position on the virtual interactive interface namely the Interface module.
At last, Interface module interprets both the position and the area of the tracked zone of interest into actions executed by Engine module.
An overview of the complete architecture of the framework is illustrated in Figure~\ref{fig:diagramarchitecture}.

From an implementation point of view, the integration of the framework in the operating system could be achieved through two different approaches.
(i) The first approach is the Direct System Integration (DSI) engine which emulates the mouse.
This mouse is fully integrated into the surrounding environment.
The mouse could for instance control a genuine keyboard and mouse software, such as the visual keyboard.
Then Mouse module maps the position into screen coordinates and Interface module does not really exist as a separate entity.
(ii) The second approach is the Interface Control (IC) engine which allows to control one particular interface which then could interpret and execute actions.


\section{Detailed Description of the Architecture}
\label{sec:3}

\subsection{Functions to Isolate Zones of Interest Module}

FIZI module, which stands for \textit{Functions to Isolate Zones of Interest}, is the module in charge of the video segmentation part.
Its main goal is to segment and select the zones of interests in each image of the video sequence.
In our case, the zones of interest are the hands of the end user, and to communicate with the machine, the end user has to perform basic hand gestures interpreted by the framework.
The FIZI module concerns all the image processing part.
The outputs is a binary image which represents the mask of the master hand.
FIZI module first runs an initialization step which aims at learning the different parameters that characterize the empty background of the application.
Then, three following processing steps are run in parallel:
(i) Background removal to select only the relevant information, i.e., the foreground area where the person is,
(ii) Grey zone removal by deleting the awkward zones, in order to reduce the luminosity dependance,
(iii) Skin color detection and segmentation to select the hands.
With these three steps, the framework builds a mask of the different skin zones on each image which could be the hands.
A selection step enable the selection of the master hand, i.e., the hand used to control--i.e. to give orders--to the machine.

As indicated previously, a first step of FIZI aims at initializing the parameters of the FIZI module by learning the main discriminative features of the background.
This learning step consists of a machine learning algorithm similar to the one presented in~\cite{magoules:journal-auth:45,magoules:journal-auth:39} for energy applications~\cite{magoules:proceedings-auth:26,magoules:proceedings-auth:25,magoules:proceedings-auth:28,magoules:proceedings-auth:79} and allows to detect key features of the background~\cite{magoules:journal-auth:47,magoules:journal-auth:31,magoules:journal-auth:42,magoules:proceedings-auth:27}.
This enable to improve the background removal procedure.

To build the mask efficiently with respect to the need of low computational time, FIZI module runs three different images in parallel.
These three images are the same image captured by the camera but in different color spaces:
(i) an RGB image is used to remove the background using the learned parameters,
(ii) another RBG image is used to remove the grey zones,
(iii) an HSV image is used to detect and to segment the skin zones in the image.
For each of these images, the processing corresponds to a basic threshold processing.

At last, a merging step is processed to build the output meaningful resulting image.
The resulting mask is obtained by a logical AND operator between the three images.
Some morphological operations, combining erosion and dilatation operators are applied to remove the small noisy objects and to connect to neighborhood zones.
As the output of FIZI module, an image which represent the set of the hand zones of the end user is obtained.

\subsection{Tracking Module}

The objectives of the Tracking module are first, to select the zone of the image corresponding to the guiding hand using image region features.
This selection step is done by labeling the region into connected regions of the outputs of FIZI module, i.e., the binary mask of the skin zones, and by the characterization of the different regions.
For each connected region, size features such as the area and position features such as the center of gravity and the location in the global image are computed.
The different regions are sorted according to these different features.
Second the Tracking module tracks the selected region on the video sequence and updates the region features for each frame.
A lot of region tracking have been proposed in the literature, and here the tracking is done frame by frame by the described region selection process and a comparison with the previous frame.
The output of the tracking module is a image region corresponding to the guiding hand together with its characteristic features.

\subsection{Mouse Module}

\begin{figure}
  \centering
  \includegraphics[scale=0.5]{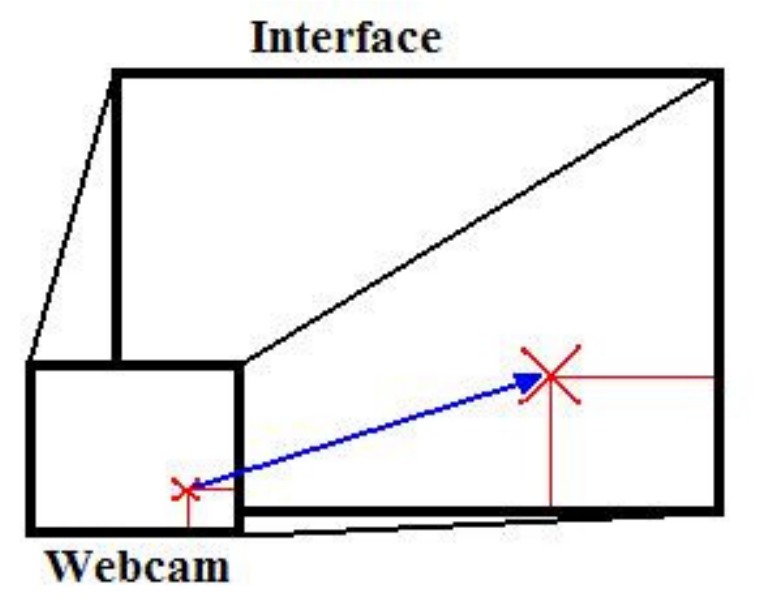}
  \includegraphics[scale=0.5]{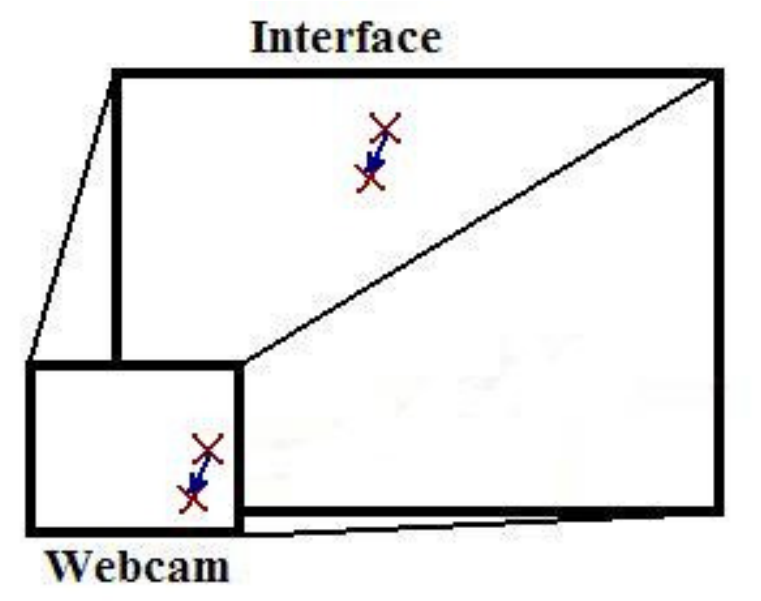}
  \includegraphics[scale=0.5]{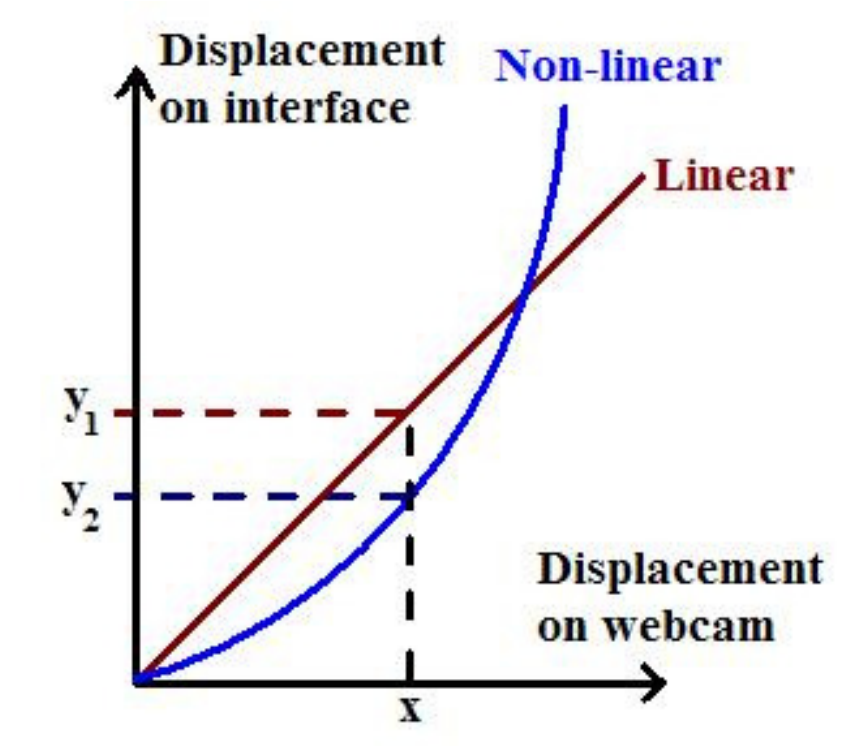}
  \caption{Diagram of the mapping approaches.
  From the left to the right, absolute, relative and relative mapping approach.}
  \label{fig:absolute}
\end{figure}

Mouse module is the module responsible for building the link between the output of the Tracking module, i.e., a zone of interest in the image corresponding to the guiding hand, with the Interface module.
The main goal of the Mouse module is the mapping of this zone of interest into a point or a cursor on the displayed interface.
This implies a strong relationship between Interface module and Mouse module.
Different mapping approaches are possible to build the link between the output of the Tracking module and the Interface module:
(i) \emph{Absolute mapping:} simple ratios are used to build the link.
This method is easy and effective when the size of the image frame buffer and the interface are quite close.
(ii) \emph{Linear relative mapping:} the relative displacements are used to move the cursor on the interface.
Quite noise sensitive, this method could be tiring for the user.
(iii) \emph{Non-linear relative mapping:} it is the same idea than the previous, but a non linear displacement function is added to ensure small displacements when the user moves its hand slowly and bigger ones when the user moves its hand faster.
This method is weakly dependent upon the noise.
Figure \ref{fig:absolute} illustrates these mapping approaches.
Absolute mapping is mainly used for Interface Control and for Direct System Integration.
For the latter, the non-linear relative mapping is more often considered.

\subsection{Interface Module}

Interface module is the module in charge of the displayed interface.
Its functionalities are:
(i) to display and to manage the virtual interface from its description in the XML language,
(ii) to control, together with the Mouse module, user interactions and associated actions.
XML language has been selected to describe these interfaces, mainly because it's user-friendly.
XML language allows the users to define the interfaces on their own.

The first step is to load and to parse the XML document.
The XML file describes the different zones of the interface and how they interact with the system.
For instance, a small square could be declared with a text label that emulate the pressing on a key, e.g., `A' when the user click on the zone.
Once the XML file has been loaded, interactions and displays are computed in order to be more efficient while processing one frame.
This approach allows to gain some precious milliseconds.
In addition, a look-up table is built to quickly determine to which zone belongs the position interacted by the user.
Once everything has been processed with the previous initialization computations, the Interface module is ready to be displayed.
This entity receives as an input the cursor's position sent by Mouse module, and compute the zone concerned by the current interaction.
Then the entity retrieves the actions to be executed and send them to the Engine module.

\subsection{Engine Module}

This last part consists of the engine in charge of the execution process.
The processed actions are sent by the Interface module to the Engine module.
Engine module is operating system dependent.
Its content is quite simple: it sends the system calls corresponding to the received orders.
The order are described by three integers: one for the action type and two as parameters.

\subsection{Implementation}

Our framework is developed in C++ with the Open-MP library and the MPI library for the parallel computations.
The image processing kernel of the framework is based on OpenCV, since it offers function for easy image manipulation and processing.
CBlob library is used for the labeling for the resulting zones after FIZI processing.
TinyXml library is used to easily and efficiently parse and load the interface description from an XML file.
The code is written in an object oriented approach.

\section{Experimental Results}
\label{sec:4}

\begin{figure}
  \centering
  \includegraphics[scale=0.17]{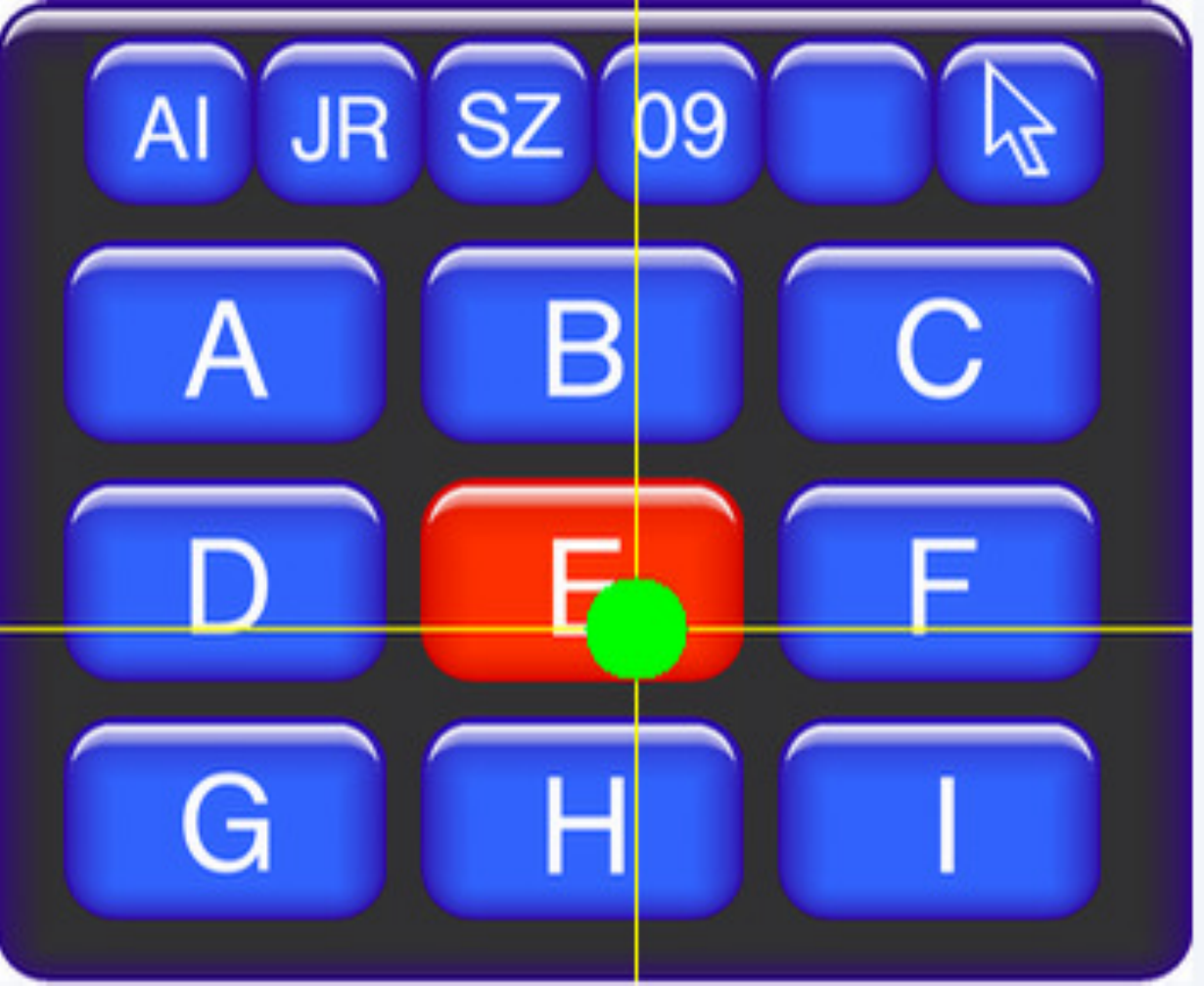}
  \includegraphics[scale=0.17]{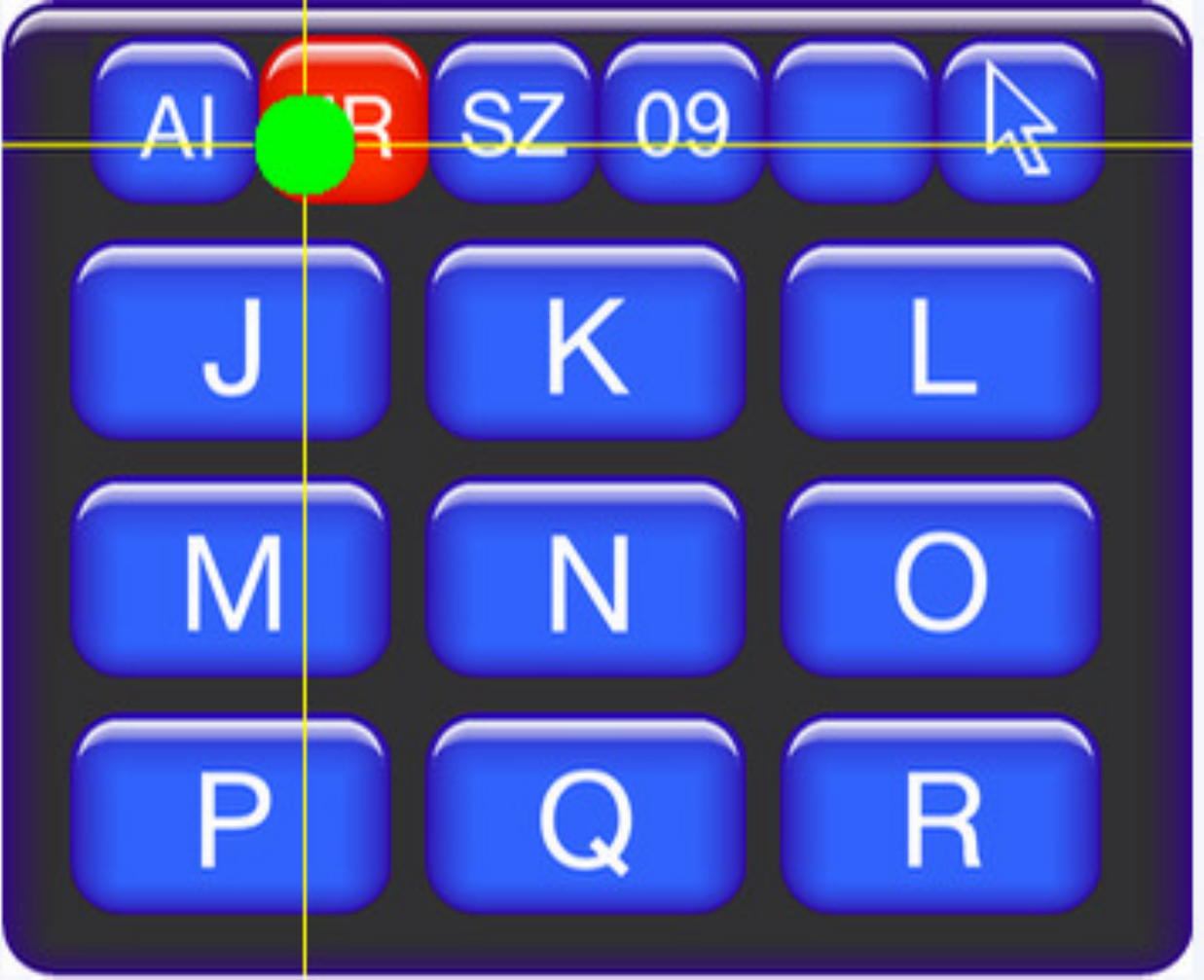}
  \includegraphics[scale=0.17]{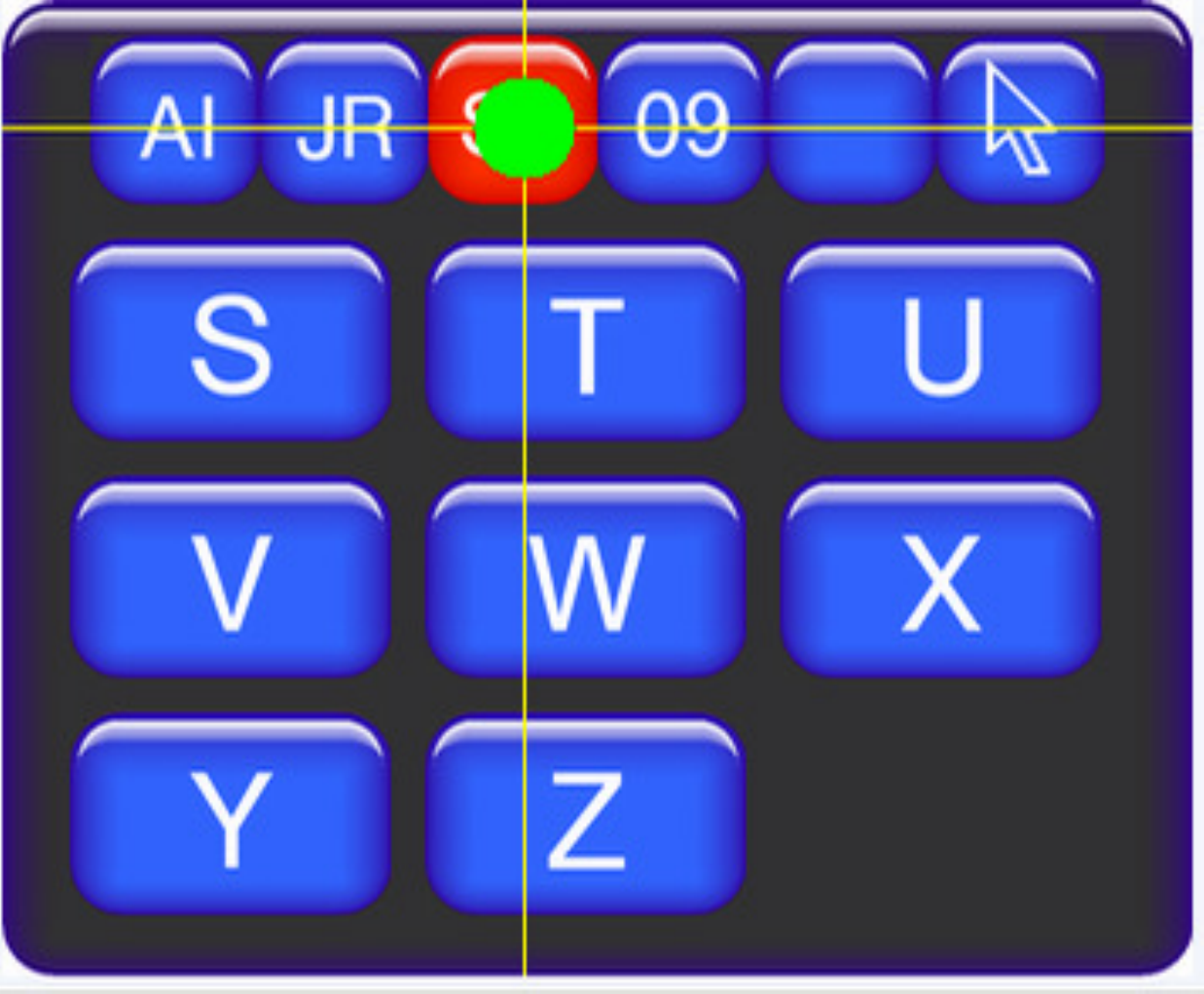}
  \includegraphics[scale=0.17]{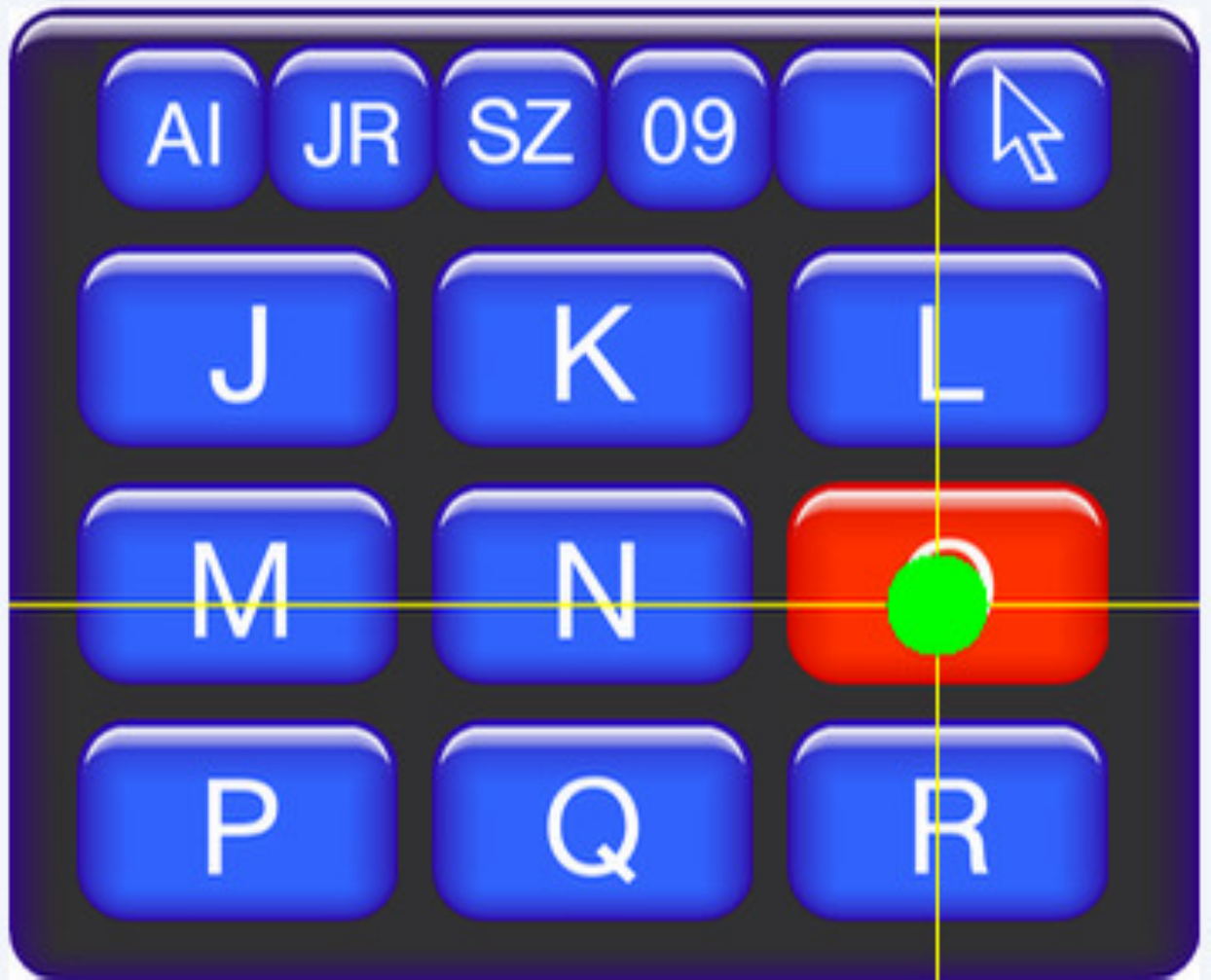}
  \includegraphics[scale=0.17]{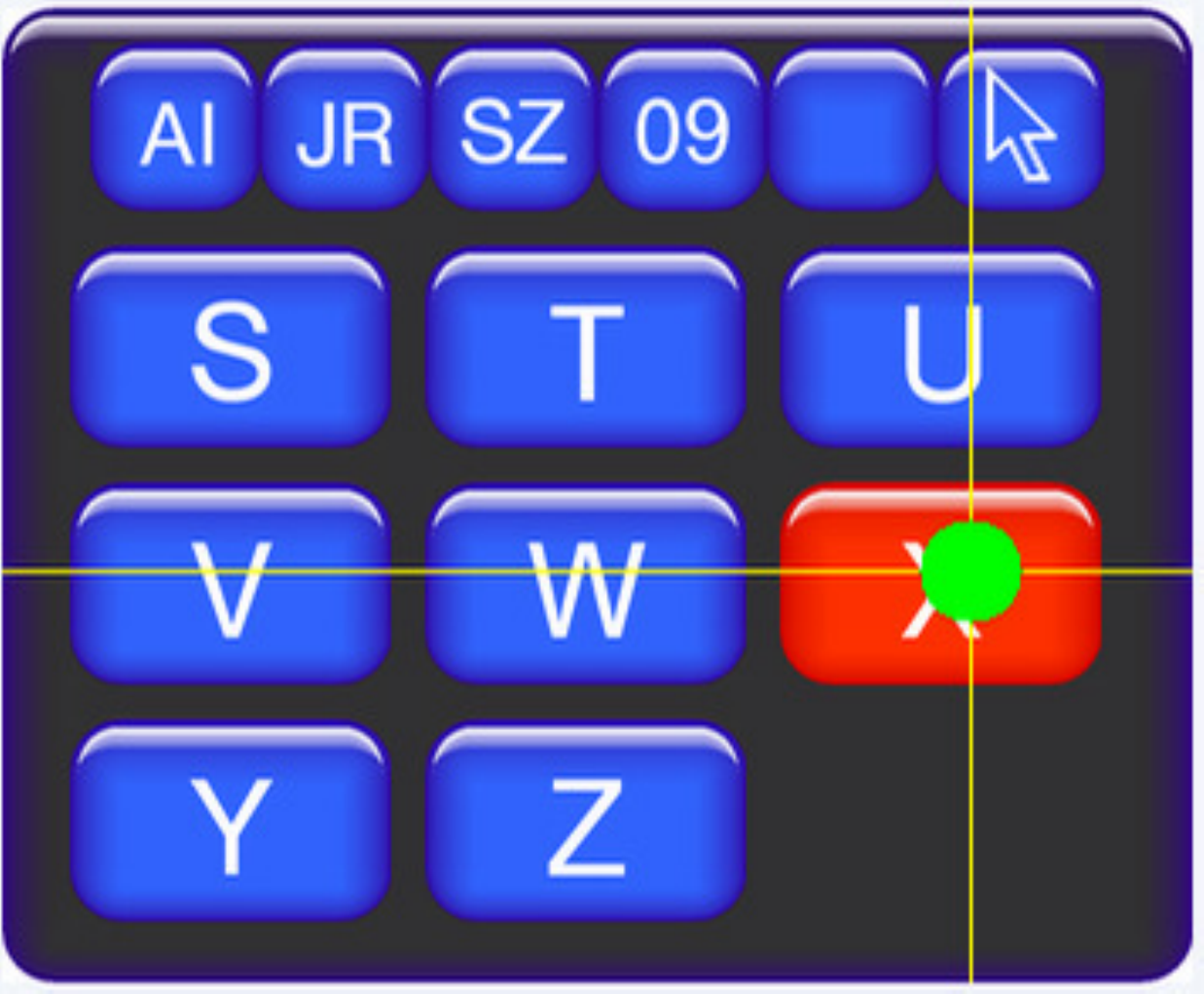}
  \includegraphics[scale=0.17]{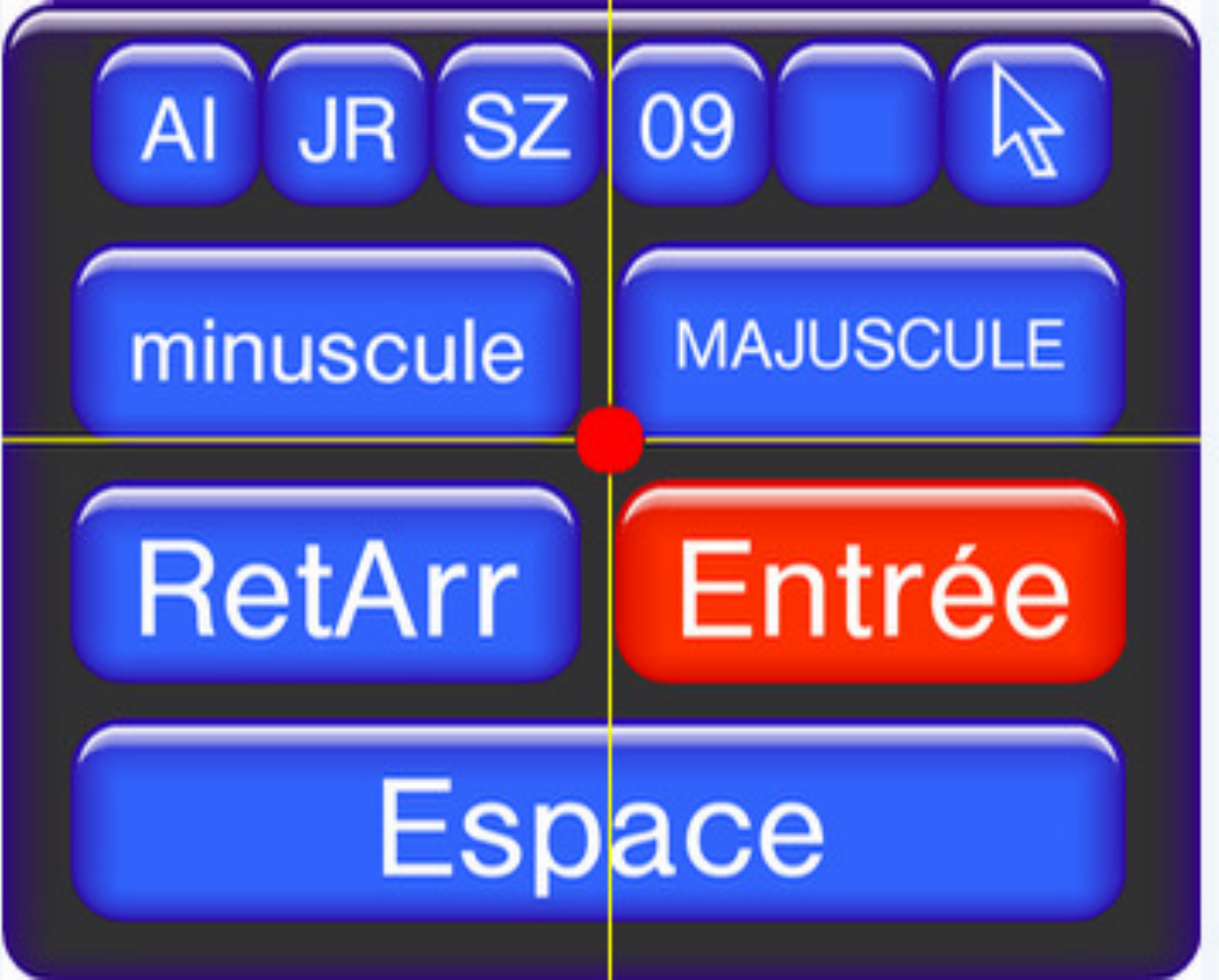}
  \caption{Sequence of gesture for typing the word `fox'.}
  \label{fig:keyboard-fox}
\end{figure}

The first interface available consists of a mouse.
All common features of a mouse are present: single left and right clic, double left clic, wheel up and down, moves.
The second interface available consists of a keyboard.
All the keys from a real keyboard are implemented: letters (A to Z), digits (0 to 9), and special keys (space, backspace, return, etc).
Of course, all the keys are not present on the same screen in order to make this interface easier to use with small displacement of the user hand.
To select a letter, the right page has to be chosen by the user.
Figure~\ref{fig:keyboard-fox} illustrates the sequence of gesture for typing the word `fox'.

\section{Conclusions}
\label{sec:5}

In this paper an original approach for contactless human interface is presented.
The proposed approach, based on computer vision and machine learning techniques, achieves a virtual mouse and a virtual keyboard using an image acquisition device.
Machine learning allows high quality of the captured images, and a parallel implementation ensures fast processing of the captured images.
This allows real time interaction of the user with the computer, without physical contact, as required for surgery applications for instance.

\section*{Acknowledgements}

The authors acknowledge the numerous students from Ecole Sup\'erieure des Sciences et Technologies de l'Ing\'enieur de Nancy (France) and from Ecole Centrale Paris (France) who have contributed to this framework since 2003, and in particular N. Vienne, J. Tavernier, the main programmers of the image processing workflow, J. Petin, N. Lambolez, M. Hjalmars, S. Cagnon, C. Mombereau, J. Ott, G. Mathias, S. Massot, D. Miliche, W. Ken, P. R\'emi, B. Rochet, J. Holburn, G. Sauwala, A. Brito Alves da Silva, A. Ortega, A. Vinicius Gonzalves Cardoso, F. Mirieu, P. d'Herbemont, E. de Roux, the main programmers of several modules, H.-X. Zhao the main programmer of the machine learning techniques.
The authors acknowledge L. Cabaret and C. Hudelot for the usefull discussions during this long term project.
Since 2006, this framework has been named ViKi (Virtual Interactive Keyboard Interface).

\bibliography{ref}

\begin{thebibliography}{10}

\bibitem{cootes95}
T.~Ahmad, C.~Taylor, A.~Lanitis, and T.~Cootes.
\newblock Tracking and recognising hand gestures using statistical shape
  models.
\newblock In {\em Proceedings of 6th British Conf on Machine vision, Vol.2},
  pages 403--412, Surrey, UK, 1995. BMVA Press.

\bibitem{cipolla1998cvh}
R.~Cipolla and A.~Pentland.
\newblock {\em {Computer vision for human machine interaction}}.
\newblock Cambridge University Press, 1998.

\bibitem{dix2004hci}
A.~Dix, J.~Finlay, G.~Abowd, and R.~Beale.
\newblock {\em {Human computer interaction}}.
\newblock Pearson Prentice Hall, 2004.

\bibitem{GiDa2005.1}
F.~Gianni and P.~Dalle.
\newblock {Interaction visuo-gestuelle avec un mur d'images}.
\newblock In {\em {Proceedings of 2nd International Society for Gesture
  Studies: Interacting Bodies / Corps en interaction , Lyon, 15-18 Jun. 2005}}.
  Ecole Normale Sup\'erieure Lettres et Sciences Humaines, juin 2005.

\bibitem{jr99survey}
J.~Joseph and J.~LaViola.
\newblock A survey of hand posture and gesture recognition techniques and
  technology.
\newblock Technical Report CS-99-11, 1999.
\newblock {Brown University Providence, RI, USA}.

\bibitem{icvs03}
R.~Kjeldsen, A.~Levas, and C.~Pinhanez.
\newblock Dynamically reconfigurable vision-based user interfaces.
\newblock {\em Mach. Vision Appl.}, 16(1):6--12, 2004.

\bibitem{magoules:journal-auth:31}
F.~Lai, F.~Magoul\`es, and F.~Lherminier.
\newblock Vapnik's learning theory applied to energy consumption forecasts in
  residential buildings.
\newblock {\em International Journal of Computer Mathematics},
  85(10):1563--1588, 2008.

\bibitem{len02}
S.~Lenmann, L.~Bretzner, and B.~Thuresson.
\newblock Computer vision based hand gesture interfaces for human computer
  interaction.
\newblock Technical report, Royal Institute of Technology of Sweden, 2002.

\bibitem{magoules:proceedings-auth:79}
F.~Magoul\`es, M.~Piliougine, and D.~Elizondo.
\newblock Support vector regression for electricity consumption prediction in a
  building in japan.
\newblock In {\em Proceedings of IEEE Intl Conf on Computational Science and
  Engineering (CSE) and IEEE Intl Conf on Embedded and Ubiquitous Computing
  (EUC) and 15th Intl Symp on Distributed Computing and Applications for
  Business Engineering (DCABES)}, pages 189--196. {IEEE CPS}, 2016.

\bibitem{magoules:journal-auth:47}
F.~Magoul\`es, H.-X. Zhao, and D.~Elizondo.
\newblock Development of an {RDP} neural network for building energy
  consumption fault detection diagnosis.
\newblock {\em Energy and Buildings}, 62:133--138, 2013.

\bibitem{martin97}
J.~Martin and J.~Crowley.
\newblock An appearance based approach to gesture-recognition.
\newblock In {\em Proceedings of 9th Intl Conf on Image Analysis and
  Processing, Vol.2}, pages 340--347, London, UK, 1997. Springer-Verlag.

\bibitem{moeslund06}
T.~Moeslund, A.~Hilton, and V.~Kruger.
\newblock A survey of advances in vision-based human motion capture and
  analysis.
\newblock {\em Computer Vision and Image Understanding}, 104(2):90--126, 2006.

\bibitem{ouhaddi-hand}
H.~Ouhaddi and P.~Horain.
\newblock 3d hand gesture tracking by model registration.
\newblock Available online at:
  {\url{citeseer.ist.psu.edu/article/ouhaddi99hand.html}} (accessed November
  2007).

\bibitem{Poppe662:2007}
R.~Poppe.
\newblock Vision based human motion analysis: an overview.
\newblock {\em Computer Vision and Image Understanding}, 108(1-2):4--18, 2007.

\bibitem{sturman1994sgb}
D.~Sturman, D.~Zeltzer, and P.~Medialab.
\newblock {A survey of glove-based input}.
\newblock {\em Computer Graphics and Applications, IEEE}, 14(1):30--39, 1994.

\bibitem{utsumi96}
A.~Utsumi, T.~Miyasato, F.~Kishino, and R.~Nakatsu.
\newblock Hand gesture recognition system using multiple cameras.
\newblock In {\em Proceedings of Intl Conf on Pattern Recognition, Vol.1}, page
  667, Washington, DC, USA, 1996. IEEE CPS.

\bibitem{wu99visionbased}
Y.~Wu and T.~Huang.
\newblock Vision based gesture recognition: a review.
\newblock {\em Lecture Notes in Computer Science}, 1739:103+, 1999.

\bibitem{magoules:proceedings-auth:25}
H.-X. Zhao and F.~Magoul\`es.
\newblock A new parallel implementation of {SVM} on multi-core systems.
\newblock In Y.~Li, editor, {\em Proceedings of Intl Conf on Modeling,
  Simulation and Control (ICMSC 2010), Cairo, Egypt, 2-4 Nov. 2010}. ISBN/ISSN:
  978-1-4244-8823-0, 2010.

\bibitem{magoules:journal-auth:39}
H.-X. Zhao and F.~Magoul\`es.
\newblock Parallel support vector machines applied to the prediction of
  multiple buildings energy consumption.
\newblock {\em Journal of Algorithms and Computational Technology},
  4(2):231--250, 2010.

\bibitem{magoules:proceedings-auth:26}
H.-X. Zhao and F.~Magoul\`es.
\newblock Feature selection for support vector regression in the application of
  building energy prediction.
\newblock In {\em Proceedings of 9th IEEE Intl Symp on Applied Machine
  Intelligence and Informatics (SAMI 2011), Smolenice, Slovakia, 27-29 Jan.
  2011}. {IEEE CPS}, 2011.

\bibitem{magoules:proceedings-auth:28}
H.-X. Zhao and F.~Magoul\`es.
\newblock New parallel support vector regression for predicting building energy
  consumption.
\newblock In {\em Proceedings of IEEE Symp Series on Computational Intelligence
  in Multicriteria Decision Making, Paris, France, April 11--15, 2011}. {IEEE
  CPS}, 2011.

\bibitem{magoules:proceedings-auth:27}
H.-X. Zhao and F.~Magoul\`es.
\newblock Parallel support vector machines on multi-core and multiprocessor
  systems.
\newblock In R.~Fox, editor, {\em Proceedings of 11th Intl Conference on
  Artificial Intelligence and Applications (AIA 2011), Innsbruck, Austria,
  February 14--16, 2011}. IASTED, 2011.

\bibitem{magoules:journal-auth:42}
H.-X. Zhao and F.~Magoul\`es.
\newblock Feature selection for predicting building energy consumption based on
  statistical learning method.
\newblock {\em Journal of Algorithms and Computational Technology},
  6(1):59--78, 2012.

\bibitem{magoules:journal-auth:45}
H.-X. Zhao and F.~Magoul\`es.
\newblock A review on the prediction of building energy consumption.
\newblock {\em Renewable and Sustainable Energy Reviews}, 16(6):3586--3592,
  2012.

\end{thebibliography}
\bibliographystyle{abbrv}

\end{document}